\documentclass[aps,prl,twocolumn,superscriptaddress,showpacs,preprintnumbers]{revtex4-1}

\usepackage{amsmath}
\usepackage{amssymb}
\usepackage{graphicx}
\usepackage{epsfig}
\usepackage{color}
\usepackage{multirow}
\usepackage{rjwmath}
\usepackage[mathcal]{eucal}
\usepackage{array}
\usepackage{booktabs}
\usepackage{wasysym}
\usepackage{chngcntr}
\usepackage{float}
\usepackage{capt-of}
\usepackage{pbox}
\begin{document}

\title{ \quad\\[1.0cm] First observation of the hadronic transition $ \Upsilon(4S) \to \eta h_{b}(1P)$ and new measurement of the $h_b(1P)$ and $\eta_b(1S)$ parameters}

\begin{abstract}
Using a sample of $771.6 \times 10^{6}$  $\Upsilon(4S)$ decays collected by the Belle experiment at the KEKB $e^+e^-$ collider, we observe for the first time the transition $\Upsilon(4S) \to \eta h_b(1P)$ with the branching fraction ${\cal B}[\Upsilon(4S) \to \eta h_b(1P)]= (2.18 \pm 0.11 \pm 0.18) \times 10^{-3}$ and we measure the $h_b(1P)$ mass $M_{h_{b}(1P)} = (9899.3 \pm 0.4 \pm 1.0)$ MeV/$c^{2}$, corresponding to the hyperfine splitting $\Delta M_{\mathrm HF}(1P) = (0.6 \pm 0.4 \pm 1.0)$ MeV/$c^{2}$.
Using the transition $h_b(1P) \to \gamma \eta_b(1S)$, we measure the $\eta_b(1S)$ mass $M_{\eta_{b}(1S)} = (9400.7 \pm 1.7 \pm 1.6)$ MeV/$c^{2}$, corresponding to $\Delta M_{\mathrm HF}(1S) = (59.6 \pm 1.7 \pm 1.6)$ MeV/$c^{2}$, the $\eta_b(1S)$ width $\Gamma_{\eta_{b}(1S)} = (8 ^{+6}_{-5} \pm 5)$ MeV/$c^{2}$ and  the branching fraction ${\cal B}[h_b(1P) \to \gamma \eta_b(1S)]= (56 \pm 8 \pm 4) \%$.
\end{abstract}

% 14.40.Pq Heavy quarkonia
% 112.38.Qk QCD	Experimental tests
% 12.39.Hg Heavy quark effective theory
% 13.20.Gd Decays of J/ψ, Υ, and other quarkonia

\noaffiliation
\affiliation{University of the Basque Country UPV/EHU, 48080 Bilbao}
\affiliation{Beihang University, Beijing 100191}
\affiliation{University of Bonn, 53115 Bonn}
\affiliation{Budker Institute of Nuclear Physics SB RAS and Novosibirsk State University, Novosibirsk 630090}
\affiliation{Faculty of Mathematics and Physics, Charles University, 121 16 Prague}
\affiliation{University of Cincinnati, Cincinnati, Ohio 45221}
\affiliation{Deutsches Elektronen--Synchrotron, 22607 Hamburg}
\affiliation{Justus-Liebig-Universit\"at Gie\ss{}en, 35392 Gie\ss{}en}
\affiliation{SOKENDAI (The Graduate University for Advanced Studies), Hayama 240-0193}
\affiliation{Gyeongsang National University, Chinju 660-701}
\affiliation{Hanyang University, Seoul 133-791}
\affiliation{University of Hawaii, Honolulu, Hawaii 96822}
\affiliation{High Energy Accelerator Research Organization (KEK), Tsukuba 305-0801}
\affiliation{IKERBASQUE, Basque Foundation for Science, 48013 Bilbao}
\affiliation{Indian Institute of Technology Guwahati, Assam 781039}
\affiliation{Indian Institute of Technology Madras, Chennai 600036}
\affiliation{Indiana University, Bloomington, Indiana 47408}
\affiliation{Institute of High Energy Physics, Vienna 1050}
\affiliation{Institute for High Energy Physics, Protvino 142281}
\affiliation{INFN - Sezione di Torino, 10125 Torino}
\affiliation{Institute for Theoretical and Experimental Physics, Moscow 117218}
\affiliation{J. Stefan Institute, 1000 Ljubljana}
\affiliation{Kanagawa University, Yokohama 221-8686}
\affiliation{Kennesaw State University, Kennesaw GA 30144}
\affiliation{King Abdulaziz City for Science and Technology, Riyadh 11442}
\affiliation{Korea Institute of Science and Technology Information, Daejeon 305-806}
\affiliation{Korea University, Seoul 136-713}
\affiliation{Kyungpook National University, Daegu 702-701}
\affiliation{\'Ecole Polytechnique F\'ed\'erale de Lausanne (EPFL), Lausanne 1015}
\affiliation{Faculty of Mathematics and Physics, University of Ljubljana, 1000 Ljubljana}
\affiliation{Luther College, Decorah, Iowa 52101}
\affiliation{University of Maribor, 2000 Maribor}
\affiliation{Max-Planck-Institut f\"ur Physik, 80805 M\"unchen}
\affiliation{School of Physics, University of Melbourne, Victoria 3010}
\affiliation{Middle East Technical University, 06531 Ankara}
\affiliation{Moscow Physical Engineering Institute, Moscow 115409}
\affiliation{Moscow Institute of Physics and Technology, Moscow Region 141700}
\affiliation{Graduate School of Science, Nagoya University, Nagoya 464-8602}
\affiliation{Kobayashi-Maskawa Institute, Nagoya University, Nagoya 464-8602}
\affiliation{Nara Women's University, Nara 630-8506}
\affiliation{National Central University, Chung-li 32054}
\affiliation{Department of Physics, National Taiwan University, Taipei 10617}
\affiliation{H. Niewodniczanski Institute of Nuclear Physics, Krakow 31-342}
\affiliation{Niigata University, Niigata 950-2181}
\affiliation{Osaka City University, Osaka 558-8585}
\affiliation{Pacific Northwest National Laboratory, Richland, Washington 99352}
\affiliation{Peking University, Beijing 100871}
\affiliation{University of Pittsburgh, Pittsburgh, Pennsylvania 15260}
\affiliation{University of Science and Technology of China, Hefei 230026}
\affiliation{Seoul National University, Seoul 151-742}
\affiliation{Soongsil University, Seoul 156-743}
\affiliation{University of South Carolina, Columbia, South Carolina 29208}
\affiliation{Sungkyunkwan University, Suwon 440-746}
\affiliation{School of Physics, University of Sydney, NSW 2006}
\affiliation{Department of Physics, Faculty of Science, University of Tabuk, Tabuk 71451}
\affiliation{Tata Institute of Fundamental Research, Mumbai 400005}
\affiliation{Excellence Cluster Universe, Technische Universit\"at M\"unchen, 85748 Garching}
\affiliation{Toho University, Funabashi 274-8510}
\affiliation{Tohoku University, Sendai 980-8578}
\affiliation{Department of Physics, University of Tokyo, Tokyo 113-0033}
\affiliation{Tokyo Institute of Technology, Tokyo 152-8550}
\affiliation{Tokyo Metropolitan University, Tokyo 192-0397}
\affiliation{University of Torino, 10124 Torino}
\affiliation{CNP, Virginia Polytechnic Institute and State University, Blacksburg, Virginia 24061}
\affiliation{Wayne State University, Detroit, Michigan 48202}
\affiliation{Yamagata University, Yamagata 990-8560}
\affiliation{Yonsei University, Seoul 120-749}
  \author{U.~Tamponi}\affiliation{INFN - Sezione di Torino, 10125 Torino}\affiliation{University of Torino, 10124 Torino} % Torino
  \author{R.~Mussa}\affiliation{INFN - Sezione di Torino, 10125 Torino} % Torino	
  \author{A.~Abdesselam}\affiliation{Department of Physics, Faculty of Science, University of Tabuk, Tabuk 71451} % Tabuk
  \author{H.~Aihara}\affiliation{Department of Physics, University of Tokyo, Tokyo 113-0033} % Tokyo
  \author{K.~Arinstein}\affiliation{Budker Institute of Nuclear Physics SB RAS and Novosibirsk State University, Novosibirsk 630090} % BINP
  \author{D.~M.~Asner}\affiliation{Pacific Northwest National Laboratory, Richland, Washington 99352} % PNNL
  \author{H.~Atmacan}\affiliation{Middle East Technical University, 06531 Ankara} % METU
  \author{T.~Aushev}\affiliation{Moscow Institute of Physics and Technology, Moscow Region 141700}\affiliation{Institute for Theoretical and Experimental Physics, Moscow 117218} % ITEP
  \author{R.~Ayad}\affiliation{Department of Physics, Faculty of Science, University of Tabuk, Tabuk 71451} % Tabuk
  \author{I.~Badhrees}\affiliation{Department of Physics, Faculty of Science, University of Tabuk, Tabuk 71451}\affiliation{King Abdulaziz City for Science and Technology, Riyadh 11442} % Tabuk
% ...
% Ljubljana
\collaboration{The Belle Collaboration}

\maketitle

%%%% >>>> keep the final version single-spaced
\tighten

{\renewcommand{\thefootnote}{\fnsymbol{footnote}}}
\setcounter{footnote}{0}

The bottomonium system, comprising bound states of $b$ and  $\bar{b}$ quarks, has been studied extensively in the past \cite{Brambilla:2010cs, Brambilla:2014}.
The recent observations of unexpected hadronic transitions from the $J^{PC} = 1^{--}$ states above the 
$B\bar{B}$ meson threshold, $\Upsilon(4S)$ and $\Upsilon(5S)$, to lower mass bottomonia 
have opened new pathways to the elusive spin-singlet states, the $h_b(nP)$ and $\eta_b(nS)$ \cite{Adachi:2012,Mizuk:2012pb}, and challenged theoretical descriptions, showing a large violation of the selection rules that apply to transitions below the threshold.

Hadronic transitions between the lowest mass quarkonium levels can be described using the QCD multipole expansion (ME)~\cite{Gottfried:1977gp, Bhanot:1979af, Peskin:1979va, Bhanot:1979vb,  Voloshin:1978hc, Voloshin:1980zf}. In this approach, the heavy quarks
emit two gluons that subsequently transform into light hadrons. The $\pi\pi$ and $\eta$ transitions between the vector states proceed via emission of E1E1 and E1M2 gluons, respectively. Therefore, $\eta$ transitions are highly suppressed as they require a spin flip of the heavy quark \cite{Kuang:2006me, Voloshin:2007dx}.
Indeed, the ratio of branching fractions
\[
\mathcal{R}^{\eta S}_{\pi\pi S}(n,m)=
\frac{\mathcal{B}[\Upsilon(nS)\to\eta\Upsilon(mS)]}
{\mathcal{B}[\Upsilon(nS)\to\pi^+\pi^-\Upsilon(mS)]}
\]
is measured to be small for low-lying states:
$\mathcal{R}^{\eta S}_{\pi\pi S}(2,1)=
(1.64\pm0.23)\times10^{-3}$~\cite{He:2008xk, BABAR:2011ab, Tamponi:2012rw} and
$\mathcal{R}^{\eta S}_{\pi\pi S}(3,1)<2.3\times10^{-3}$~\cite{BABAR:2011ab}.

Above the $B\bar{B}$ threshold, BaBar observed the transition
$\Upsilon(4S)\to\eta\Upsilon(1S)$ with the unexpectedly large branching fraction of
$(1.96\pm0.28)\times10^{-4}$, corresponding  to
$\mathcal{R}^{\eta S}_{\pi\pi S}(4,1)=2.41\pm0.42$~\cite{Aubert:2008az}.
This apparent violation of the heavy quark spin-symmetry was explained
by the contribution of $B$ meson loops or, equivalently, by the
presence of a four-quark $B\bar{B}$ component inside the $\Upsilon(4S)$
wave function~\cite{Meng:2008,Voloshin:2011}.
At the $\Upsilon(5S)$ energy, the anomaly is even more striking. The spin-flip processes $\Upsilon(5S)\to \pi\pi h_b(1P, 2P)$ are found not to be suppressed with respect to  the spin-symmetry  preserving reactions $\Upsilon(5S)\to \pi\pi \Upsilon(1S, 2S)$ \cite{Adachi:2012}, and all the $\pi\pi$ transitions show the presence of new resonant structures \cite{Bondar:2012,Krokovny:2013} that cannot be explained as conventional bottomonium states. 

Further insight into the mechanism of the hadronic transitions above the threshold can be gained by
searching for the $E1M1$ transition $\Upsilon(4S)\to\eta h_b(1P)$,
which is  predicted to have a branching fraction of the order of
$10^{-3}$~\cite{Guo:2010ca}.

In this Letter, we report the first observation of the
$\Upsilon(4S)\to\eta h_b(1P)$ transition and measurement of the
$h_b(1P)$ and $\eta_b(1S)$ resonance parameters. 
Following the approach used for the observation of the $h_b(1P, 2P)$ production in $e^+e^-$ collisions at the $\Upsilon(5S)$ energy \cite{Adachi:2012} --- by studying the 
inclusive $\pi^+\pi^-$ missing mass in hadronic events --- we investigate the missing mass spectrum of $\eta$ mesons in the $\Upsilon(4S)$ data sample. The missing mass is defined as
$M_{\rm miss}(\eta)=\sqrt{(P_{e^+e^-}-P_{\eta})^2}$, where $P_{e^+e^-}$
and $P_{\eta}$ are the four-momenta of the colliding
$e^+e^-$  pair and the $\eta$ meson, respectively.

The large sample of reconstructed $h_b(1P)$ events allows us to measure its mass and,
via the $h_b(1P)\to\gamma\eta_b(1S)$ transition, the mass and width
of the $\eta_b(1S)$. The latter are especially important since there is
a $3.2\,\sigma$ discrepancy between the $\eta_b(1S)$ mass measurement
by Belle using $h_b(1P,2P)\to\gamma\eta_b(1S)$
transitions~\cite{Mizuk:2012pb} and by BaBar and CLEO using
$\Upsilon(2S,3S)\to\gamma\eta_b(1S)$
~\cite{Aubert:2008ba,Aubert:2009as,Bonvicini:2009hs}.

This analysis is based on
the $711$ fb$^{-1}$ sample collected at the centre-of-mass energy of $\sqrt{s} = 10.580$ GeV/$c^2$ by the Belle experiment \cite{Abashian:2000cg, Brodzicka:2012} at the KEKB asymmetric-energy $e^+ e^-$ collider \cite{kekb,kekb_1,kekb_2}, corresponding to $771.6\times10^6$ $\Upsilon(4S)$ decays.
Monte Carlo (MC) samples are generated using EvtGen \cite{Lange:2001uf}. The detector response is simulated with GEANT3 \cite{Geant3}. 
Separate MC samples are generated for each run period to account for the changing detector performance and accelerator conditions.

Candidate events are requested to satisfy the standard Belle hadronic selection \cite{Abe:2001hj}, to have at least three charged tracks pointing towards the primary interaction vertex, a visible energy greater than $0.2\sqrt{s}$, a total energy deposition in the electromagnetic calorimeter (ECL) between $0.1\sqrt{s}$ and $0.8\sqrt{s}$, and a total momentum balanced along the $z$ axis. Continuum $e^+e^- \to q\bar{q}$ events (where $q \in \{u,\, d,\, s,\, c\}$) are suppressed by requiring $R_{2}$, the ratio of the $2^\mathrm{nd}$ to $0^\mathrm{th}$ Fox-Wolfram moment \cite{Fox:1978}, to be less than $0.3$.
The $\eta$ candidates are reconstructed in the dominant $\eta \to \gamma \gamma$ channel. 
The $\gamma$ candidates are selected from energy deposits in the ECL that have a shape compatible with an electromagnetic shower, and are not associated with charged tracks.
We investigate the absolute photon energy calibration using three calibration samples: $\pi^0 \to \gamma \gamma$, $\eta \to \gamma\gamma$, and $D^{*0} \to D^{0}\gamma$ \cite{Mizuk:2012pb}. Comparing the peak position and the widths of the three calibration signals in the MC sample and in the data, as a function of the photon energy $E$, we determine the photon energy correction ${\cal F}_{en}(E)$ and the resolution fudge factor ${\cal F}_{res}(E)$. We observe ${\cal F}_{en}(E) < 0.1\%$ and  ${\cal F}_{res}(E) \approx (+5 \pm 3)\%$ in the signal region, and apply the corresponding correction to the MC samples.
An energy threshold, ranging from  $50$ MeV to $95$ MeV, is applied as a function of the polar angle to reject low energy photons arising from the beam-related backgrounds.  
To reject photons from $\pi^0$ decays, $\gamma\gamma$ pairs having invariant mass within $17$ MeV/$c^2$ of the nominal $\pi^0$ mass \cite{PDG:2012} are identified as $\pi^0$ candidates and the corresponding photons are excluded from the $\eta$ reconstruction process.  
The angle $\theta$ between the photon direction and that of the $\Upsilon(4S)$ in the $\eta$ rest frame peaks at $\cos(\theta) \approx 1$ for the remaining combinatorial background. We thus require  $\cos(\theta) < 0.94$ for the $\eta$ selection. 
All the selection criteria are optimized using the MC simulation by maximizing the figure of merit $f = N_\mathrm{sig}/\sqrt{N_\mathrm{sig} + N_\mathrm{bkg}}$, where $N_\mathrm{sig}$ and $N_\mathrm{bkg}$ are the signal and background yields in the signal region, respectively.
\begin{figure*}%%%%%%%%%%%%%%%
%\begin{center}
   \includegraphics[width=1.\linewidth]{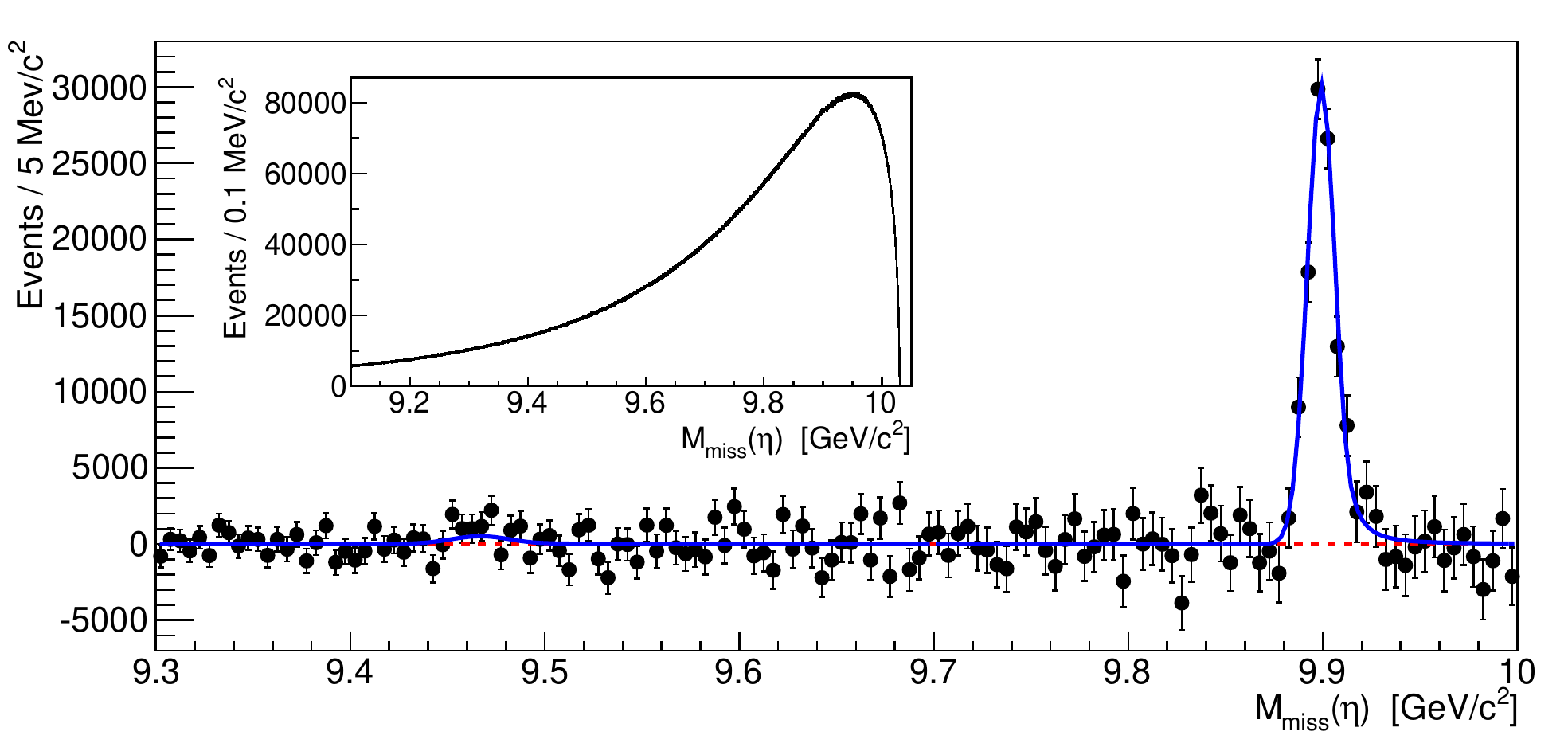}
   \caption{$M_{\mathrm{miss}}(\eta)$ distribution after the background  subtraction. The solid blue curve shows the fit with the signal PDFs, while the dashed red curve represents the background only hypothesis. The inset shows the $M_{\mathrm{miss}}(\eta)$ distribution before the background subtraction.}
   \label{fig:4S_hb_Data}
%\end{center}
\end{figure*}%%%%%%%%%%%%%%%
The $\eta$ peak in the $\gamma\gamma$ invariant mass distribution, after the selection is applied, can be fit by a Crystal Ball (CB) \cite{Gaiser:1986} probability density function (PDF) with a resolution of $13$ MeV/$c^2$. Thus, $\gamma\gamma$ pairs with an invariant mass  within $26$ MeV/$c^2$ of the nominal $\eta$ mass $m_{\eta}$ \cite{PDG:2012} are selected as a signal sample, while the candidates in the regions $39$ MeV/$c^2$ $<|M(\gamma\gamma) - m_{\eta}| < 52$ MeV/$c^2$ are used as control samples. 
To improve the $M_{\mathrm{miss}}(\eta)$ resolution, a mass-constrained fit is performed on the $\eta$ candidates in both the signal and control regions.
The resulting $M_{\mathrm{miss}}(\eta)$ distribution is shown in the inset of Fig. \ref{fig:4S_hb_Data}.
The $\Upsilon(4S) \to \eta h_b(1P)$ and $\Upsilon(4S) \to \eta \Upsilon(1S)$ peaks in $M_{\mathrm{miss}}(\eta)$ are modeled with a CB PDF, whose Gaussian core resolutions are fixed according to the MC simulation. The parameters of the non-Gaussian tails, which account for the effects of the soft Initial State Radiation (ISR), are calculated assuming the next-to-leading order formula for the ISR emission probability \cite{Benayoun:1999} and by modeling the $\Upsilon(4S)$ as a Breit-Wigner resonance with $\Gamma = (20.5 \pm 2.5)$ MeV/$c^2$ \cite{PDG:2012}. 
The $M_{\mathrm{miss}}(\eta)$ spectrum is fitted in two separate intervals: $(9.30, 9.70)$ GeV/$c^2$ and $(9.70, 10.00)$ GeV/$c^2$. In the first (second) interval, the combinatorial background is described with a $6^\mathrm{th}$-order ($11^\mathrm{th}$) Chebyshev polynomial. The polynomial order is determined maximizing the credibility level of the fit and is validated using the sideband samples.
Figure \ref{fig:4S_hb_Data} shows the background-subtracted $M_{\mathrm{miss}}(\eta)$ distribution, with a bin size 50 times larger than that used for the fit. 
The credibility levels of the fits are $1\%$ in the lower interval and $19\%$ in the upper one.
The transition $\Upsilon(4S) \to \eta h_b(1P)$ is observed with a statistical significance of $11 \sigma$, calculated using the profile likelihood method \cite{Cowan:2010}, and no signal is observed in the $\gamma\gamma$-mass control regions. The $h_b(1P)$ yield is $N_{h_b(1P)} = 112469 \pm 5537$. From the position of the peak, we measure $M_{h_{b}(1P)} = (9899.3 \pm 0.4 \pm 1.0)$ MeV/$c^{2}$ (hereinafter the first error is statistical and the second is systematic). 
We calculate the branching fraction of the transition as
$$
{\cal B}[\Upsilon(4S) \to \eta h_b(1P)]  =   \frac{N_{h_b(1P)}}{N_{\Upsilon(4S)} \epsilon_{\eta h_b(1P)} {\cal B}[\eta \to \gamma\gamma]},
$$
where $N_{\Upsilon(4S)} = (771.6 \pm 10.6)\times 10^{6}$ is the number of $\Upsilon(4S)$, $\epsilon_{\eta h_b(1P)} = (16.96 \pm 1.12)\%$ is the reconstruction efficiency and ${\cal B}[\eta \to \gamma\gamma] = (39.41 \pm 0.21)\%$ \cite{PDG:2012}. We obtain ${\cal B}[\Upsilon(4S) \to \eta h_b(1P)] = (2.18 \pm 0.11 \pm 0.18 ) \times 10^{-3}$, in agreement with the available theoretical prediction \cite{Guo:2010ca}. No evidence of $\Upsilon(4S) \to \eta \Upsilon(1S)$ is present, so we set the 90$\%$ Credibility Level (CL) upper limit ${\cal B}[\Upsilon(4S) \to \eta \Upsilon(1S)] < 2.7 \times 10^{-4}$, in agreement with the previous experimental result by BaBar \cite{Aubert:2008az}. All the upper limits presented in this work are obtained using the $CL_{s}$ technique \cite{Junk:1999,Read:2002} and include systematic uncertainties.  
Using our measurement of $M_{h_b(1P)}$, we calculate the corresponding $1P$ hyperfine splitting, defined as the difference between the $\chi_{bJ}(1P)$ spin-averaged mass $m^{\mathrm sa}_{\chi_{bJ}(1P)}$  and the $h_b(1P)$ mass, and obtain $\Delta M_{\mathrm HF}(1P) = (+ 0.6 \pm 0.4 \pm 1.0)$ MeV/$c^{2}$; the systematic error includes the  uncertainty on the  value of $m^{\mathrm sa}_{\chi_{bJ}(1P)}$ \cite{PDG:2012}.

As validation of our measurement, we study the $\eta \to \pi^+\pi^-\pi^0$ mode. The $\pi^0$ candidate is reconstructed from a $\gamma\gamma$ pair with invariant mass within $17$ MeV/$c^2$ of the nominal $\pi^0$ mass \cite{PDG:2012} while the $\pi^{\pm}$ candidates tracks are required to be associated with the primary interaction vertex and not identified as kaons by the particle identification algorithm. We observe an excess in the signal region with statistical significance of 3.5$\sigma$ and measure
${\cal B}[\Upsilon(4S) \to \eta h_b(1P)]_{\eta \to \pi^+\pi^-\pi^0}= (2.3 \pm 0.6) \times 10^{-3}$, which is in agreement with the result from the $\gamma\gamma$ mode. 

The contributions to the systematic uncertainty in our measurements are summarized in Table \ref{tab:syst4Shb}. To estimate them, we first vary --- simultaneously --- the fit ranges within $\pm 100$ MeV/$c^2$ and the order of the background polynomial between 7 (4) and 14 (8) in the upper (lower) interval. The average variation of the fitted parameters when the fitting conditions are so changed is adopted as the fit-range/model systematic uncertainty.   Similarly, we vary the bin width between $0.1$ and $1$ MeV/$c^2$ and we treat the corresponding average variations as the bin-width systematic error. The ISR modeling contribution is due to the $\Upsilon(4S)$ width uncertainty \cite{PDG:2012}. 
The presence of peaking backgrounds is studied using MC samples of inclusive $B\bar{B}$ events and bottomonium transitions.
While no peaking background due to $B$ meson decay has been identified, the as-yet-unobserved transitions $\Upsilon(4S) \to \gamma\gamma\Upsilon(1^3D_{1,2}) \to \gamma\gamma\eta \Upsilon(1S)$ can appear in the $M_{\mathrm{miss}}(\eta)$ spectrum as a CB-shaped peaking structure at $M_{\mathrm{miss}}(\eta) = 9.877$ GeV/$c^2$ with resolution of $10.6$ MeV/$c^2$. We take this effect into account by repeating the fit with and without an additional CB component. No signal is observed and we obtain the upper limit on the product of branching fractions ${\cal B}[\Upsilon(4S) \to \gamma\gamma\Upsilon(1^3D_{1,2})]\times{\cal B}[\Upsilon(1^3D_{1,2}) \to \eta \Upsilon(1S)] < 0.8 \times 10^{-4}$ ($90\%$ CL). 
The uncertainty on the photon energy calibration factors is determined by varying both ${\cal F}_{en}(E)$ and ${\cal F}_{res}(E)$ within their errors.
The uncertainty on the reconstruction efficiency includes contributions from several sources. Using $121.4$ fb$^{-1}$ collected at the $\Upsilon(5S)$ energy, the $\Upsilon(5S) \to \pi^+\pi^- \Upsilon(2S)$ transition is reconstructed; comparing the $R_{2}$ shape obtained from this data sample with the simulation provides a $\pm 3\%$ uncertainty related to the continuum rejection. A $\pm 1 \%$ uncertainty is assigned for the efficiency of the hadronic event selection. 
\begin{table}
\small\caption{Systematic uncertainties  in the determination of ${\cal B}[\Upsilon(4S) \to \eta h_b(1P)]$, in units of \%, and on $M_{h_b(1P)}$, in units of MeV/$c^{2}$. }
    \label{tab:syst4Shb}
%\begin{center}
\begin{tabular}{lcc} 
\hline Source                                  & ${\cal B}$  &   $M_{h_b(1P)}$ \\
\hline 
Fit range and background PDF order             &  $\pm$2.4 &   $\pm$0.1   \\
Bin width                                      &  $\pm$2.5 &   $\pm$0.1   \\
ISR modeling                                   &  $\pm$2.8 &   $\pm$0.7   \\
Peaking backgrounds                            &  $\pm$0.5 &   $\pm$0.4   \\
$\gamma$ energy calibration                    &  $\pm$1.2 &   $\pm$0.3   \\
Reconstruction efficiency                      &  $\pm$6.6 &     -        \\
$N_{\Upsilon(4S)}$                             &  $\pm$1.4 &     -        \\
Beam energy                                    &  $\pm$0.0 &   $\pm$0.4   \\
${\cal B}[\eta \to \gamma\gamma]$              &  $\pm$0.5 &     -        \\
\hline
Total                                          &  $\pm$8.2 &   $\pm$1.0   \\
\hline
\end{tabular}
%\end{center}
\end{table} 
The uncertainty on the photon reconstruction efficiency is estimated using $D \to K^{\pm}\pi^{\mp}\pi^{0}$ events to be $\pm 2.8\%$ per photon, corresponding to $\pm 5.6\%$ per $\eta$. The number of $\Upsilon(4S)$ mesons is measured with a relative uncertainty of $\pm 1.4 \%$ from the number of hadronic events after the subtraction of the continuum contribution using off-resonance data. 
The absolute value of accelerator beam energies are calibrated by fully reconstructed $B$ mesons. We observe a $\pm 0.4$
MeV/c$^{2}$ fluctuation of $M_{h_b(1P)}$ due to the uncertainty on the $B$ meson mass \cite{PDG:2012} and a negligible effect on the branching ratio measurement.
Finally, we include an uncertainty in the branching fraction due to the uncertainty in ${\cal B}[\eta \to \gamma \gamma]$ \cite{PDG:2012}.

\begin{figure}[h!]%%%%%%%%%%%%%%%
%\begin{center}
   \includegraphics[width=1.\linewidth]{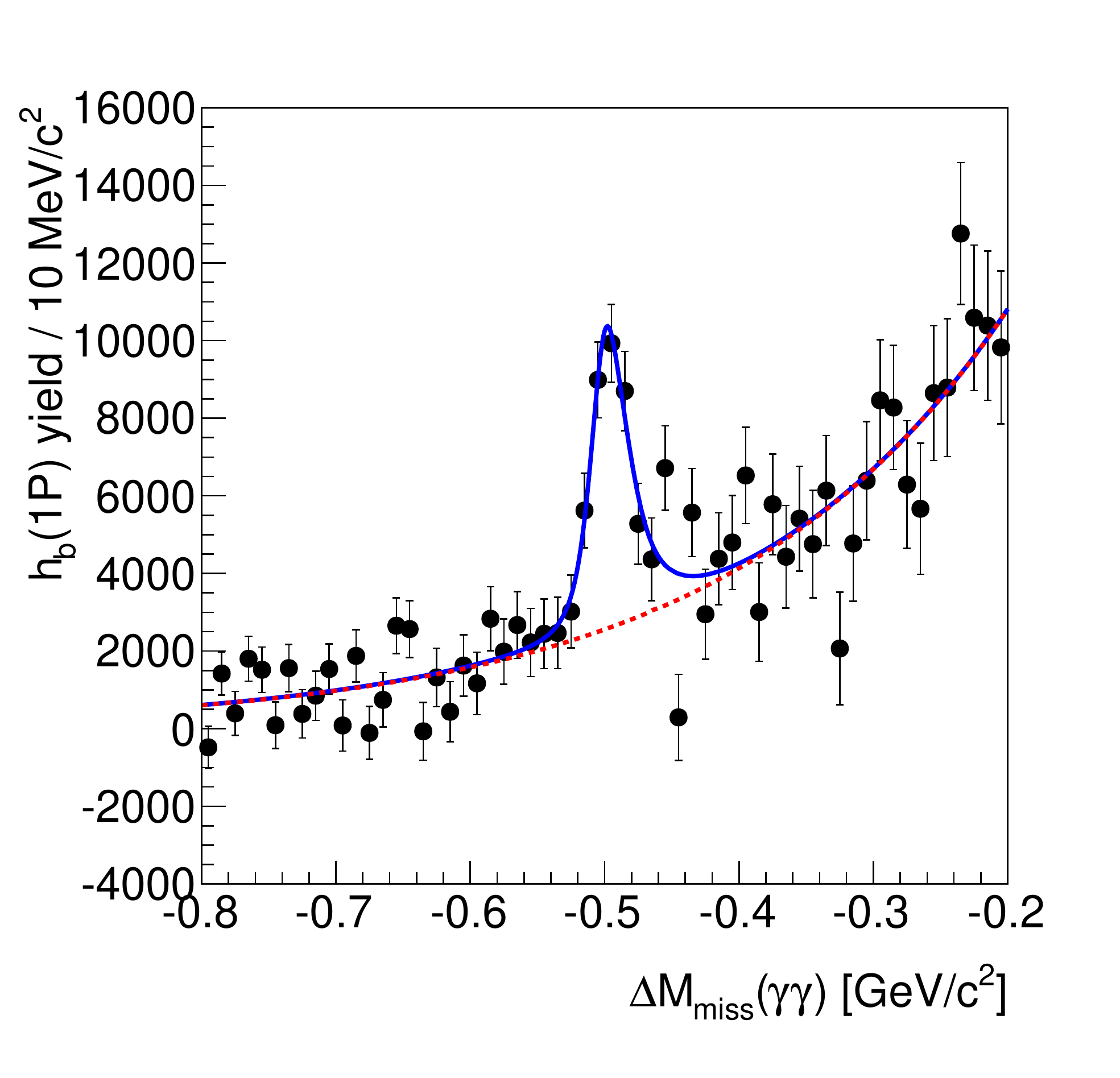}
   \caption{$\Delta M_{\mathrm{miss}}$ distribution. The blue solid line shows our best fit, while the red, dashed line represents the background component. }
   \label{fig:DeltaMMdata}
%\end{center}
\end{figure}

The study of the $\eta_b(1S)$ is performed by reconstructing the transitions $\Upsilon(4S) \to \eta h_b(1P) \to \eta \gamma \eta_b(1S)$. To extract the signal, we measure the number of $\Upsilon(4S) \to \eta h_b(1P)$ events $N_{h_b(1P)}$ as a function of the variable  $\Delta M_{\mathrm{miss}} = M_{\mathrm{miss}}(\eta\gamma) - M_{\mathrm{miss}}(\eta)$, where $M_{\mathrm{miss}}(\eta\gamma)$ is the missing mass of the $\eta\gamma$ system. The signal transition will produce a peak in $N_{h_b(1P)}$ at $m_{\eta_b(1S)} - m_{h_b(1P)}$.
The radiative photon arising from the $h_b(1P)$ decay is reconstructed with the same criteria used in the $\eta \to \gamma\gamma$ selection, and the $h_b(1P)$ yield in each $\Delta M_{\mathrm{miss}}$ bin is measured with the  fitting procedure described above. 
To assure the convergence of the $M_{\mathrm miss}(\eta)$ fit in each $\Delta M_{\mathrm miss}$ interval, the $h_b(1P)$ mass is fixed to $9899.3$ MeV/$c^2$, the range is reduced to  $(9.80, 9.95)$ GeV/$c^2$ and the order of the background PDF polynomial is decreased to seven. The $h_b(1P)$ yield as function of $\Delta M_{\mathrm{miss}}$, shown in Fig. \ref{fig:DeltaMMdata}, exhibits an excess at $\Delta M_{\mathrm{miss}} = M_{\eta_b(1S)}- M_{h_b(1P)}$ with a statistical significance of 9$\sigma$. 
%%%%%%%%%%%%%%%
The $\eta_b(1S)$ peak is described by the convolution of a double-sided CB PDF, whose parameters are fixed according to the MC simulation, and a non-relativistic Breit-Wigner PDF that accounts for the natural $\eta_b(1S)$ width. The background is described by an exponential. 
We measure $M_{\eta_b(1S)}- M_{h_b(1P)} = (-498.6 \pm 1.7 \pm 1.2)$ MeV/$c^{2}$,  $\Gamma_{\eta_b(1S)} = (8 ^{+6}_{-5} \pm 5)$ MeV/$c^2$ and the number of $\Upsilon(4S) \to \eta h_b(1P) \to \eta \gamma \eta_b(1S)$ events  $N_{\eta_b(1S)} = 33116 \pm 4741$. The credibility level of the fit is $50\%$.
We calculate the branching fraction of the radiative transition as  
$$
{\cal B}[h_b(1P) \to \gamma \eta_b(1S)]  = \frac{N_{\eta_b(1S)} \epsilon_{\eta h_b(1P)}}{N_{h_b(1P)} \epsilon_{\eta \gamma \eta_b(1S)}}, 
$$
where $\frac{\epsilon_{\eta h_b(1P)}}{\epsilon_{\eta \gamma \eta_b(1S)}} = 1.887 \pm 0.053$ is the ratio of the reconstruction efficiencies  for $\Upsilon(4S) \to  \eta h_b(1P)$ and $\Upsilon(4S) \to  \eta h_b(1P) \to \eta \gamma \eta_b(1S)$. We obtain ${\cal B}[h_b(1P) \to \gamma \eta_b(1S)]= (56 \pm 8 \pm 4) \%$.
To estimate the systematic uncertainties reported in Table \ref{tab:systEtabmass},  we adopt the protocols discussed earlier.
\begin{table}[h!]
\small
    \caption{Systematic uncertainties in the determination of the $\eta_b(1S)$ mass and width, in units of MeV/$c^{2}$ and on ${\cal B} = {\cal B}[h_b(1P) \to \gamma \eta_b(1S)$ , in units of \%.}
    \label{tab:systEtabmass}
%\begin{center}
\begin{tabular}{lccc} 
\hline Source                                  & $\Delta M_{\mathrm{miss}}$    &   $\Gamma_{\eta_b(1S)}$ &  ${\cal B}$ \\
\hline     
$M_{\mathrm{miss}}(\eta)$ fit range            &  $\pm$0.8            &   $\pm$3.0              &  $\pm$2.8 \\
$M_{\mathrm{miss}}(\eta)$ bin width            &  $\pm$0.0            &   $\pm$0.1              &  $\pm$0.0\\
$M_{\mathrm{miss}}(\eta)$ polynomial order     &  $\pm$0.1            &   $\pm$1.9              &  $\pm$1.6 \\
$M_{h_b(1P)}$                                  &  $\pm$0.0            &   $\pm$0.8              &  $\pm$1.1\\
$\Delta M_{\mathrm{miss}}$ fit range           &  $\pm$0.0            &   $\pm$0.7              &  $\pm$2.2 \\
$\Delta M_{\mathrm{miss}}$ bin width           &  $\pm$0.8            &   $\pm$2.8              &  $\pm$5.2\\
$\gamma$ energy calibration                    &  $\pm$0.5            &   $\pm$0.3              &  $\pm$1.2 \\
Reconstruction efficiency ratio                &       -              &          -              &  $\pm$2.8 \\
\hline
Total                                          &  $\pm$1.2            &   $\pm$4.7              &  $\pm$7.2  \\
\hline
\end{tabular}
%\end{center}
\end{table}
Uncertainties related to the $M_{\mathrm{miss}}(\eta)$ fit are determined by changing the fit range, the bin width, the background-polynomial order and the fixed values of $M_{h_b(1P)}$ used in the fits. Similarly, the uncertainties arising from the $\Delta M_{\mathrm{miss}}$ fit are studied by repeating it with different ranges and binning. The calibration uncertainty accounts for the errors on the photon energy calibration factors. The uncertainty due to the ratio of the reconstruction efficiencies arises entirely from the single-photon reconstruction efficiency. The $\eta_{b}(1S)$ annihilates into two gluons, while the $h_b(1P)$ annihilates predominantly into three gluons, but the MC simulation indicates no significant difference in the $R_{2}$ shape. Therefore, the continuum suppression cut does not contribute to the uncertainty arising from the reconstruction efficiency ratio.
We calculate the $\eta_b(1S)$ mass as $M_{\eta_b(1S)} = M_{h_b(1P)} + \Delta M_{\mathrm{miss}} = (9400.7 \pm 1.7 \pm 1.6)$ MeV/$c^2$. Assuming $m_{\Upsilon(1S)} = (9460.30 \pm 0.26)$ MeV/$c^2$ \cite{PDG:2012}, we calculate $\Delta M_{\mathrm HF}(1S) = (59.6 \pm 1.7 \pm 1.6)$ MeV/$c^{2}$.

A summary of the results presented in this work is shown in Table \ref{tab:summary}.
We report the first observation of a single-meson transition from spin-triplet to spin-singlet  bottomonium states,  $\Upsilon(4S) \to \eta h_b(1P)$. This process is found to be the strongest known transition from the $\Upsilon(4S)$ meson to lower bottomonium states. 
A new measurement of the $h_b(1P)$ mass is presented. The corresponding $1P$ hyperfine splitting is compatible with zero, which can be interpreted as evidence of the absence of sizable long range spin-spin interactions.
Exploiting the radiative transition $h_b(1P) \to \gamma \eta_b(1S)$, we present a new measurement of the mass difference between the $h_b(1P)$ and the $\eta_b(1S)$ and, assuming our measurement of $M_{h_b(1P)}$, we calculate $M_{\eta_b(1S)}$. Our result is in agreement with the value obtained with the $\Upsilon(5S) \to \pi^+\pi^- h_b(1P) \to \pi^+\pi^- \gamma \eta_b(1S)$ process \cite{Mizuk:2012pb} but exhibits a discrepancy with the M1-based measurements \cite{Aubert:2008ba,Aubert:2009as,Bonvicini:2009hs}. From the theoretical point of view, our result is in agreement with the predictions of many potential models and lattice calculations \cite{Burns:2012pc}, including the recent lattice result in Ref. \cite{Dowall:2014}.
Our measurement of ${\cal B}[h_b(1P) \to \gamma \eta_b(1S)]$ agrees with the theoretical predictions \cite{Godfrey:2002, Chen:2013cpa}.
All the direct measurements presented in this work are independent of the previous results reported by Belle \cite{Adachi:2012}, which were obtained by reconstructing different transitions and using a different data sample. Furthermore, all the results except for $\Delta M_{\mathrm HF}(1S)$ and $\Delta M_{\mathrm HF}(1P)$ are obtained within the analysis described herein and are uncorrelated with the existing world averages.
\begin{table}[h!]
\small
    \caption{Summary of the results of the searches for $\Upsilon(4S) \to \eta h_b(1P)$ and $h_b(1P) \to \gamma \eta_b(1S)$.}
    \label{tab:summary}
%\begin{center}
\begin{tabular}{lc} 
\hline Observable                                 & Value  \\
\hline       
${\cal B}[\Upsilon(4S) \to \eta h_b(1P)]$         & $(2.18 \pm 0.11 \pm 0.18) \times 10^{-3}$ \\
${\cal B}[h_b(1P) \to \gamma \eta_b(1S)]$         & $(56 \pm 8 \pm 4) \%$ \\ 
$M_{h_{b}(1P)}$                                   & $(9899.3 \pm 0.4 \pm 1.0)$ MeV/$c^{2}$ \\
$M_{\eta_{b}(1S)} - M_{h_{b}(1P)}$                & $(-498.6 \pm 1.7 \pm 1.2)$ MeV/$c^{2}$ \\
$\Gamma_{\eta_{b}(1S)}$                           & $(8^{+6}_{-5} \pm 5)$ MeV/$c^{2}$ \\
$M_{\eta_{b}(1S)}$                                & $(9400.7 \pm 1.7 \pm 1.6)$ MeV/$c^{2}$ \\ 
\hline
$\Delta M_{\mathrm HF}(1S)$                       & $(+59.6 \pm 1.7 \pm 1.6)$ MeV/$c^{2}$ \\ 
$\Delta M_{\mathrm HF}(1P)$                       & $(+0.6 \pm 0.4 \pm 1.0)$ MeV/$c^{2}$ \\ 
\hline
\end{tabular}
%\end{center}
\end{table}

%***** Acknowledgments *****
We thank the KEKB group for excellent operation of the
accelerator; the KEK cryogenics group for efficient solenoid
operations; and the KEK computer group, the NII, and 
PNNL/EMSL for valuable computing and SINET4 network support.  
We acknowledge support from MEXT, JSPS and Nagoya's TLPRC (Japan);
ARC and DIISR (Australia); FWF (Austria); NSFC (China); MSMT (Czechia);
CZF, DFG, and VS (Germany); DST (India); INFN (Italy); 
MOE, MSIP, NRF, GSDC of KISTI, and BK21Plus (Korea);
MNiSW and NCN (Poland); MES (particularly under Contract No. 14.A12.31.0006), RFAAE and RFBR 
under Grant No. 14-02-01220 (Russia); ARRS (Slovenia);
IKERBASQUE and UPV/EHU (Spain); 
SNSF (Switzerland); NSC and MOE (Taiwan); and DOE and NSF (USA).

%***** Acknowledgments *****

\end{document}